\documentclass[fleqn,10pt]{wlscirep}

\title{Quantum phase estimation using path-symmetric entangled states}

\author[1,2,*]{Su-Yong Lee}
\author[1]{Chang-Woo Lee}
\author[3]{Jaehak Lee}
\author[3,1]{Hyunchul Nha}
\affil[1]{School of Computational Sciences, Korea Institute for Advanced Study, Hoegi-ro 85,Dongdaemun-gu, Seoul 02455, Korea}
\affil[2]{Centre for Quantum Technologies, National University of Singapore, 3 Science Drive 2, 117543 Singapore, Singapore}
\affil[3]{Department of Physics, Texas A\&M University at Qatar, Education City, POBox 23874, Doha, Qatar}

\affil[*]{papercrane79@kias.re.kr}

\begin{abstract}
We study the sensitivity of phase estimation using a generic class of path-symmetric entangled states $|\varphi\rangle|0\rangle+|0\rangle|\varphi\rangle$, where an arbitrary state $|\varphi\rangle$ occupies one of two modes in quantum superposition.  With this generalization, we identify the fundamental limit of phase estimation under energy constraint that is characterized by the photon statistics of the component state $|\varphi\rangle$. We show that quantum Cramer-Rao bound (QCRB) can be indefinitely lowered with super-Poissonianity of the state $|\varphi\rangle$. For possible measurement schemes, we demonstrate that a full photon-counting employing the path-symmetric entangled states achieves the QCRB over the entire range $[0,2\pi]$ of unknown phase shift $\phi$ whereas a parity measurement does so in a certain confined range of $\phi$. By introducing a component state of the form $|\varphi\rangle=\sqrt{q}|1\rangle+\sqrt{1-q}|N\rangle$, we particularly show that an arbitrarily small QCRB can be achieved even with a finite energy in an ideal situation. This component state also provides the most robust resource against photon loss among considered entangled states over the range of the average input energy $N_{av}>1$. Finally we propose experimental schemes to generate these path-symmetric entangled states for phase estimation. 
\end{abstract}
\begin{document}

\flushbottom
\maketitle
%
%
\thispagestyle{empty}


\section*{Introduction}

It is a task of fundamental and practical interest to identify the ultimate precision in measuring an unknown parameter of a system. In particular, estimating an unknown phase has many important applications such as the observation of gravitational waves \cite{A16} and detection of weak signals or defects leading to the design of highly sensitive sensors \cite{Taylor13}.
Typically, an interferometric scheme, e.g. Mach-Zehnder interferometer (MZI), is employed to measure an unknown phase $\phi$ present along one arm of the interferometer and its sensitivity $\delta\phi$ is investigated under the constraint of average input energy $N_{av}$. If a coherent state is used as an input to the MZI, it is well known that the phase sensitivity is given by $\delta\phi\sim1/\sqrt{N_{av}}$, referred to as shot-noise limit. On the other hand, the phase sensitivity can be further enhanced to $\delta\phi\sim1/N_{av}$ by using quantum resources such as squeezed states and entangled states \cite{C81,GLM06} under ideal conditions. However, as photons carrying information undergo a loss mechanism in realistic situations, it becomes a practically important question what kind of resources and measurement schemes can yield phase sensitivity robust against loss.

Here we consider the problem of local quantum estimation which aims at minimizing the variance of the estimator at a fixed value of a parameter \cite{BC94}.
 It is well known that $NOON$ state provides Heisenberg limit (HL) in local quantum phase estimation \cite{D08}. There arises a question what if the single-mode component state $|N\rangle$ is replaced by an arbitrary state $|\varphi\rangle$. 
It is also known that entanglement is not a crucial resource necessary to achieve Heisenberg limit under certain conditions\cite{Tilma10}, e.g. using separable squeezed states \cite{A10}. On the other hand, it is still an interesting and important issue to identify resource states, whether entangled or separable, which can give enhanced performance desirably under practical situations.
We here address the problem of estimating an unknown phase using a general class of two-mode entangled states as a probe, specifically those states represented by $|\varphi\rangle_a|0\rangle_b+|0\rangle_a|\varphi\rangle_b$. This generalized class of states is worth investigating comprehensively for a phase-estimation problem. 
First, $NOON$ state extensively studied so far \cite{D08,GM10,GLM11,BSV15} and the entangled-coherent state $|\alpha\rangle_a|0\rangle_b+|0\rangle_a|\alpha\rangle_b$ known to give better performance than $NOON$ state all belong to this class \cite{GC01,GB02,Joo11,J12}. Second, these states are all path-symmetric states according to Ref.\cite{H09}, which means, e.g., that a parity measurement can achieve the quantum Cram\'er-Rao bound \cite{SKDL13}, opening a venue for practical applications as well.

The upper limit of quantum phase estimation is determined by quantum Fisher information (QFI) from an information-theoretic perspective \cite{BC94}. 
An arbitrarily large QFI with a finite energy is achievable, e.g., with a superposition of vacuum and squeezed states \cite{RL12} and a specific form of superpositions of $NOON$ states \cite{Zhang13}.   
It is also known that path-symmetric entangled pure states can achieve the QFI with photon counting experiments that resolve different photon numbers \cite{H09}.
As a specific measurement scheme, parity measurement was employed in trapped ion system \cite{BIWH96} and also in optical interferometry \cite{GC01,C00}. 
The parity measurement can be realized using photon number resolving detection. It may be practically demanding to observe sub shot-noise limit due to required high-level of photon number resolution. Nevertheless, there were some experimental progresses overcoming this limit e.g. using silicon photo-multiplier detector \cite{CIDE14} or homodyne detection with sign-binning process \cite{DJA13}, manifesting super-resolution of optical measurements in the shot-noise limit.
For another estimation scheme, one can use the whole photon-counting statistics on both output modes of MZI. This gives a classical Fisher information whose inverse represents phase sensitivity, and it was also employed in experiment with a small-photon-number entangled state \cite{K10}.

In this paper, we introduce a generic class of path-symmetric entangled states and investigate their QFIs to obtain the limit of phase sensitivity for each resource state. In particular, we show that it is possible to achieve an arbitrarily large QFI using this class of states even with a finite energy in an ideal situation.  An arbitrarily large QFI does not immediately lead to an infinite precision of phase measurement in practice, requiring some prior information or an arbitrariliy many repetitions of experiment \cite{HBZW12,Hall12}. Next, we consider two specific measurement schemes for phase estimation under a practical situation with loss, i.e. parity measurement at one output mode and photon-counting measurement at both output modes. 
We discuss how the phase sensitivity varies with the photon-counting statistics of the single-mode component state $|\varphi\rangle$ under two measurement schemes. 
Finally, we propose experimental schemes to generate the path-symmetric entangled states under current technology. 

\section*{Results}


\subsection*{Quantum Fisher Information of path-symmetric entangled states}

We begin by introducing a path-symmetric entangled state, 
\begin{equation}
|\psi\rangle_{ab}=\frac{1}{\sqrt{2(1+p_0)}}(|\varphi\rangle_a|0\rangle_b+|0\rangle_a|\varphi\rangle_b ),
\end{equation}
where an arbitrary single-mode state $|\varphi\rangle$ can be expressed as $|\varphi\rangle=\sum^{\infty}_{n=0}e^{i\theta_n}\sqrt{p_n}|n\rangle$ with $\sum^{\infty}_{n=0}p_n=1$ and $p_0=|\langle0|\varphi\rangle|^2$ is the overlap between the vacuum $|0\rangle$ and the state $|\varphi\rangle$.
The photon counting statistics (PCS) of a single-mode component $|\varphi\rangle$ thus becomes a crucial factor for the performance of phase estimation.
In particular, the Mandel $Q$-factor, $Q_M=\langle\varphi |\Delta\hat{n}^2|\varphi\rangle/\langle\varphi |\hat{n}|\varphi\rangle-1$, plays a dominant role in determining Quantum Fisher information (QFI) \cite{SQ15}. The PCS with $-1\leq Q_M<0$, $Q_M=0$, and $0<Q_M<\infty$ are sub-Poissonian, Poissonian, and super-Poissonian statistics, occurring e.g. with a photon number, a coherent, and a squeezed vacuum state, respectively. 
We study the connection of the Mandel $Q$ factor to the sensitivity of phase estimation.
As shown below, given a fixed amount of input energy, the QFI is enhanced with a large variance of a single-mode component in the path-symmetric entangled state. 

For a pure state, QFI is given by $F_Q=4(\langle\dot{\psi}_{out}|\dot{\psi}_{out}\rangle-|\langle\dot{\psi}_{out}|\psi_{out}\rangle|^2)$, where
$|\psi_{out}\rangle=\hat{U}_{\phi}|\psi\rangle$ and $|\dot{\psi}_{out}\rangle=\partial |\psi_{out}\rangle /\partial \phi $ with
$\hat{U}_{\phi}$ a phase shift operation \cite{BC94,BCM96}.
Whether the phase shift occurrs in only one arm or in both arms, one would expect an identical result as long as the phase difference is the same. However, Ref. \cite{JD12} pointed out that two phase-shift configurations can give different quantum Cramer-Rao bounds. This inconsistency can be resolved by considering a phase-averaged state as an input \cite{JD12}, which also makes a practical sense particularly when one has no access to external phase reference. It turns out that the QCRB for the phase-averaged state of our consideration is the same as that of a pure-state occurrung when the phase-shift operation is applied to both of the arms. 
Applying a phase shift operation to two optical paths, $|\psi_{out}\rangle=e^{-i\phi(\hat{n}_a-\hat{n}_b)/2}|\psi\rangle$, 
the QFI is given by the variance of number difference for the input two-mode state,
\begin{eqnarray}
F_Q=\langle \psi|(\hat{n}_b-\hat{n}_a)^2|\psi\rangle - \langle \psi|(\hat{n}_b-\hat{n}_a)|\psi\rangle^2, 
\end{eqnarray}
where $|\psi\rangle=|\psi\rangle_{ab}$, $\hat{n}_a=\hat{a}^{\dag}\hat{a}$ and $\hat{n}_b=\hat{b}^{\dag}\hat{b}$.  
The average input energy is given by $N_{av}=\langle \psi |(\hat{n}_a+\hat{n}_b)|\psi\rangle$.
Repeating measurements $m$ times, the QFI provides the quantum Cram\'er-Rao bound (QCRB) as $(\delta\phi_c)^2\geq 1/(mF_Q)$ \cite{BC94,ZPK10}.

The QFI of the state in Eq. (1) is characterized by the average input energy and the Mandel $Q$-factor of the single-mode component,
\begin{eqnarray}
F_Q=\frac{\langle \hat{n}\rangle}{1+p_0}(\langle \hat{n}\rangle+1+Q_M),
\end{eqnarray}
where $\langle \hat{n}\rangle=\langle\varphi |\hat{n}|\varphi\rangle$.
For a single-shot measurement with a given energy, the corresponding QCRB is thus determined by the Mandel $Q$-factor of the single-mode component. 
That is, 
{\it Observation : Given a fixed amount of input energy, quantum Cram\'er-Rao bound is lowered with $Q_M$, thus the super-Poissonian statistics of a single-mode component in the path-symmetric entangled state gives a best result.} 

For comparison, let us consider three different states for $|\varphi\rangle$, i.e. a photon number state $|N\rangle$, a coherent state $|\alpha\rangle$, and a squeezed vacuum state $|\xi\rangle$, as representing sub-Poissonian, Poissonian, and super-Poissonian statistics, respectively. Furthermore, we introduce a superposition of single photon and $N$ photon states, $|\varphi\rangle=\sqrt{q}|1\rangle+\sqrt{1-q}|N\rangle$.
This last choice is motivated by the general QFI in Eq. (3): For a fixed energy $\langle\hat{n}\rangle$, the QFI depends on two parameters $Q_M$ and $p_0$. 
We may set $p_0=0$ (no vacuum component) to optimize the QFI, which leaves only one free parameter $Q_M$. For the case of $|\varphi\rangle=\sqrt{q}|1\rangle+\sqrt{1-q}|N\rangle$ (no vacuum component), the average energy is expressed by $\langle\hat{n}\rangle=q+(1-q)N$, which in turn gives $q=\frac{N-\langle\hat{n}\rangle}{N-1}$. Therefore, we may treat $N$ as a single free parameter, which can contribute to an arbitrarily large Mandel-$Q$ factor thereby enhancing QFI, under energy-constraint $\langle\hat{n}\rangle$.
 
The first three states, $|N\rangle$, $|\alpha\rangle$, and $|\xi\rangle$, exhibit sub-Poissonian, Poissonian, and super-Poissonian statistics, respectively. 
The superposition state $\sqrt{q}|1\rangle+\sqrt{1-q}|N\rangle$ may exhibit all kinds of photon counting statistics with the parameter $N$.
For convenience, we call the corresponding entangled states as $NOON$ state, $AOOA$ state ($A$: a coherent state), $SOOS$ state ($S$: a squeezed vacuum state), and $QOOQ$ state ($Q$: superposition ratio of $|1\rangle$ and $|N\rangle$ states). 
The corresponding QFIs are readily derived as a function of the average photon number $\langle\hat{n}\rangle$, as shown in Fig. 1. 
For $QOOQ$ state, its $F_Q$ is given by $q+(1-q)N^2=(N+1)\langle\hat{n}\rangle-N$, which shows a Heisenberg-limit scaling $O(\langle\hat{n}\rangle)$ with respect to input energy $\langle\hat{n}\rangle$ while its multiplicative constant $N+1$ varies with the free parameter $N$.
Therefore, it can take an arbitrarily large value by letting the free parameter $N\rightarrow\infty$ even under any finite energy $\langle\hat{n}\rangle$.
For the other states, the ratio of $F_Q/N_{av}$ has the order $NOON<AOOA<SOOS$. 
In Fig. 1, we show the QCRB as a function of $N_{av}$. 
We illustrate the case of $QOOQ$ state with $N=8$ and $N=100$. The state with $N=8$ does not attain a better phase sensitivity than $SOOS$ state. However $QOOQ$ state with $N=100$ beats $SOOS$ state in all ranges of $N_{av}>1$. In general, the larger $N$ in $QOOQ$ state is, the wider the range of $N_{av}$ of beating all other states is.
The plot clearly shows that QCRB can be arbitrarily lowered with super-Poissonianity of the single-mode component in the path-symmetric entangled state. 

\begin{figure}
\centerline{\scalebox{0.33}{\includegraphics[angle=0]{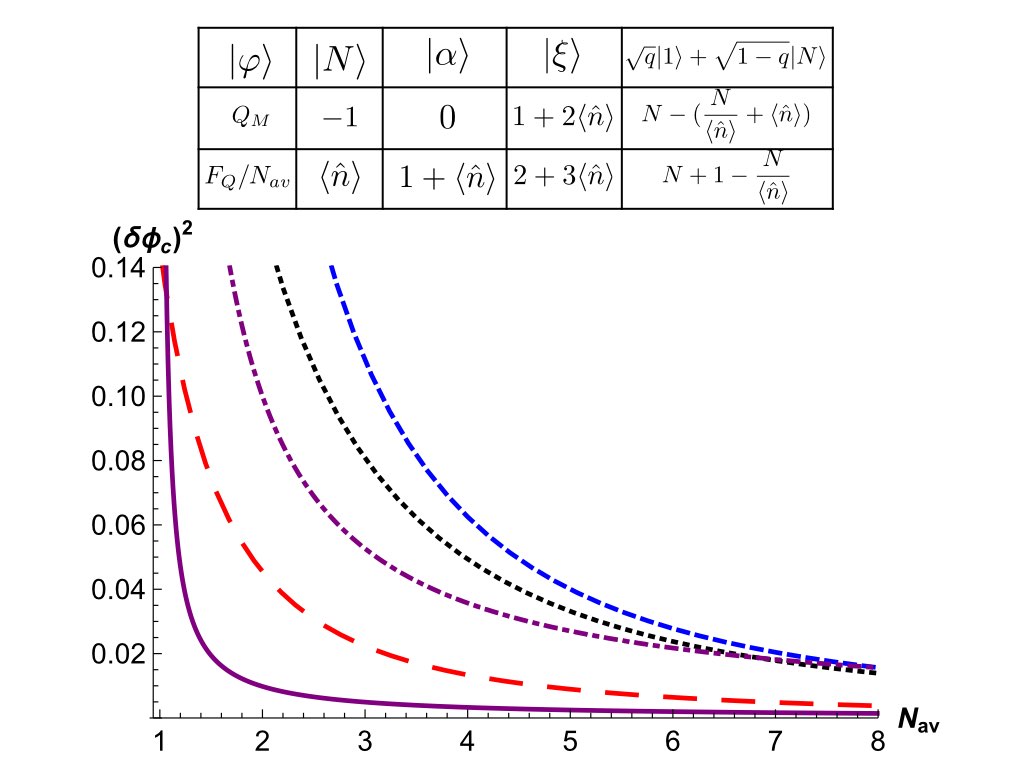}}}
\caption{Quantum Cram\'er-Rao bound as a function of $N_{av}$, using $NOON$ (blue dashed curve), $AOOA$ (black dotted curve),
$SOOS$ (red long-dashed curve), and $QOOQ$ (purple curves) states. The purple curves represent the $QOOQ$ states with $N=8$ (dot-dashed) and $100$ (solid). $A$ stands for a coherent state, $S$ for a squeezed vacuum state, and $Q$ for $|\varphi\rangle=\sqrt{q}|1\rangle+\sqrt{1-q}|N\rangle$.
}
\label{fig:fig1}
\end{figure}

{\it Phase-averaged state:} 
QFI addresses how sensitively an input state can change under the variation of the phase parameter $\phi$. It may give a different result depending on the actual configuration of phase shift, namely, whether the phase operation acts on two modes 
$e^{-i\phi(\hat{n}_a-\hat{n}_b)/2}$ or on a single mode $e^{i\phi\hat{n}_b}$, even though the phase difference $\phi$ is the same. This inconsistency may be addressed by averaging the two modes of an input state over a common phase $\theta$ with an additional reference beam \cite{JD12}.
Here, applying the phase-averaged method to the path-symmetric entangled state in Eq. (1), we obtain the phase-averaged state in a form of $NOON$-state mixtures,
\begin{eqnarray}
\rho_m=\frac{1}{1+p_0}\sum^{\infty}_{n=0}p_n|noon\rangle_{ab}\langle noon|,
\end{eqnarray}
where $|noon\rangle_{ab}=\frac{|n\rangle_a|0\rangle_b+|0\rangle_a|n\rangle_b}{\sqrt{2}}$.
Then, we find that the corresponding QFI becomes equivalent to Eq. (3) irrespective of the phase shift operations, $e^{-i\phi(\hat{n}_a-\hat{n}_b)/2}$ or $e^{i\phi\hat{n}_b}$, thus 
the above {\it Observation} also applies to the phase-averaged state.
Note that the QFI of a mixed state can be obtained by a diagonalization of the mixed state \cite{KSD11}. 


\subsection*{Measurement setup I: parity measurement on either of two output modes}

From now on, we consider some concrete measurement schemes for the phase estimation under a loss mechanism. 
As a first scheme, we consider a parity measurement in the Mach-Zehnder (MZ) interferometer in Fig. 2, of which 
the phase sensitivity is given by \cite{A10}
\begin{eqnarray}
(\Delta\phi_p)^2=\frac{1-\langle \hat{\Pi}_b\rangle^2}{(\partial\langle \hat{\Pi}_b\rangle/\partial \phi )^2}=
\frac{1-\langle \hat{\mu}_{ab}\rangle^2}{(\partial\langle \hat{\mu}_{ab}\rangle/\partial \phi )^2}. 
\end{eqnarray}
$\langle \hat{\Pi}_b\rangle=\langle(-1)^{{\hat n}_b}\rangle$ represents a parity measurement on mode $b$ after the beam splitter, which can be expressed by $\langle \hat{\mu}_{ab}\rangle=\sum^{\infty}_{L=0}\sum^L_{M=0}|L-M\rangle_a\langle M|\otimes |M\rangle_b\langle L-M|$ including all the phase-carrying off-diagonal terms in the two modes before the beam splitter.
With the path-symmetric entangled states, the resulting $\Delta\phi_p$ does not depend on the type of phase-shift operation, $\exp[i\phi\hat{n}_b]$ or $\exp[-i\frac{\phi}{2}(\hat{n}_a-\hat{n}_b)]$, before the measurements. 

The parity measurement for the state in Eq. (1) gives 
\begin{eqnarray}
\langle \hat{\mu}_{ab}\rangle=\frac{1}{1+p_0} \left[p_0+\sum^{\infty}_{n=0}p_n\cos(n\phi) \right],
\end{eqnarray}
and the corresponding phase sensitivity can be obtained by using the above expression of $\Delta\phi_p$.
In the case of $\cos(n\phi)\approx1$, the phase sensitivity is equivalent to QCRB, i.e.,
$(\Delta\phi_p)^2=1/F_Q=(\delta\phi_c)^2$, by L'H\^opital's rule.
It means that the information-theoretic optimal bound can be achieved by the path-symmetric entangled resource going through a phase-shift at $\phi\approx2\pi l/n$ ($l=0,1,2,...$).
On the other hand, notably, only $NOON$ state achieves the optimal bound for the entire range $[0, 2\pi]$ of phase shifts.

\begin{figure}
\centerline{\scalebox{0.2}{\includegraphics[angle=0]{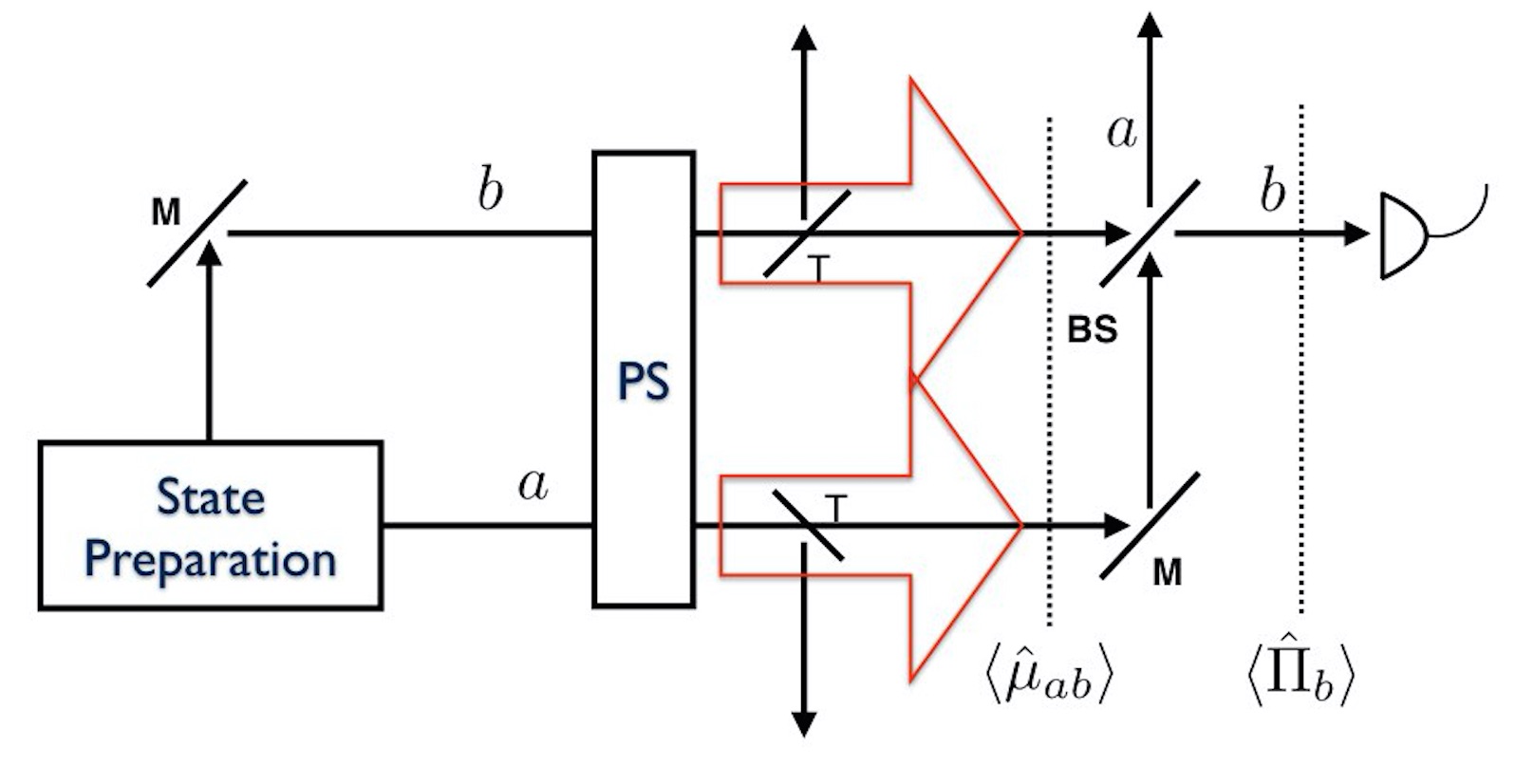}}}
\caption{Mach-Zehnder interferometer for phase estimation, for which an input state is the path-symmetric entangled state in Eq. (1).
PS stands for a phase shifter, $\exp[i\phi\hat{n}_b]$ or $\exp[-i\frac{\phi}{2}(\hat{n}_a-\hat{n}_b)]$, and M for mirror. BS is a 50:50 beam splitter.
{\it T} is the transmission rate of the beam splitters.
}
\label{fig:fig2}
\end{figure}

For a realistic situation with loss, we assume that the entangled state in Eq. (1) goes through a beam splitter acting on each mode with same transmission rate $T$ after a phase shift operation, as shown in Fig. 2. After the loss channel in the two optical paths, the parity measurement gives
\begin{eqnarray}
\langle\hat{\mu}_{ab}\rangle_T=\frac{1}{1+p_0}\sum^{\infty}_{n=0}p_n[R^n+T^n\cos(n\phi)],
\end{eqnarray}
where $R=1-T$. From the considered single-mode components $|\varphi\rangle=|N\rangle,~|\alpha\rangle,~|\xi\rangle,$ and $\sqrt{q}|1\rangle+\sqrt{1-q}|N\rangle$, we find that sub-Poissonianity of $|\varphi\rangle$ gives a better result robust against the loss channel ($T=0.9$) than (super) Poissonianity, in constrast to the QFI under an ideal situation ($T=1$). 
In Fig. 3 (a), we see that the shot-noise limit (SNL) $(\Delta\phi_p)^2_{SNL}=1/N_{av}$ is beaten in a wider range of phase-shift by $NOON$ state than the other path-symmetric states. 
It shows a trend between the photon counting statistics of $|\varphi\rangle$ and the performance of phase estimation under the loss channel. 
When the photon counting statistics changes from super-Poissonian to sub-Poissonian, the corresponding path-symmetric entangled state shows better performance, with the robustness in the order of $QOOQ(N=100)<SOOS<QOOQ(N=8)<AOOA<NOON$.


\begin{figure}
\centerline{\scalebox{0.21}{\includegraphics[angle=0]{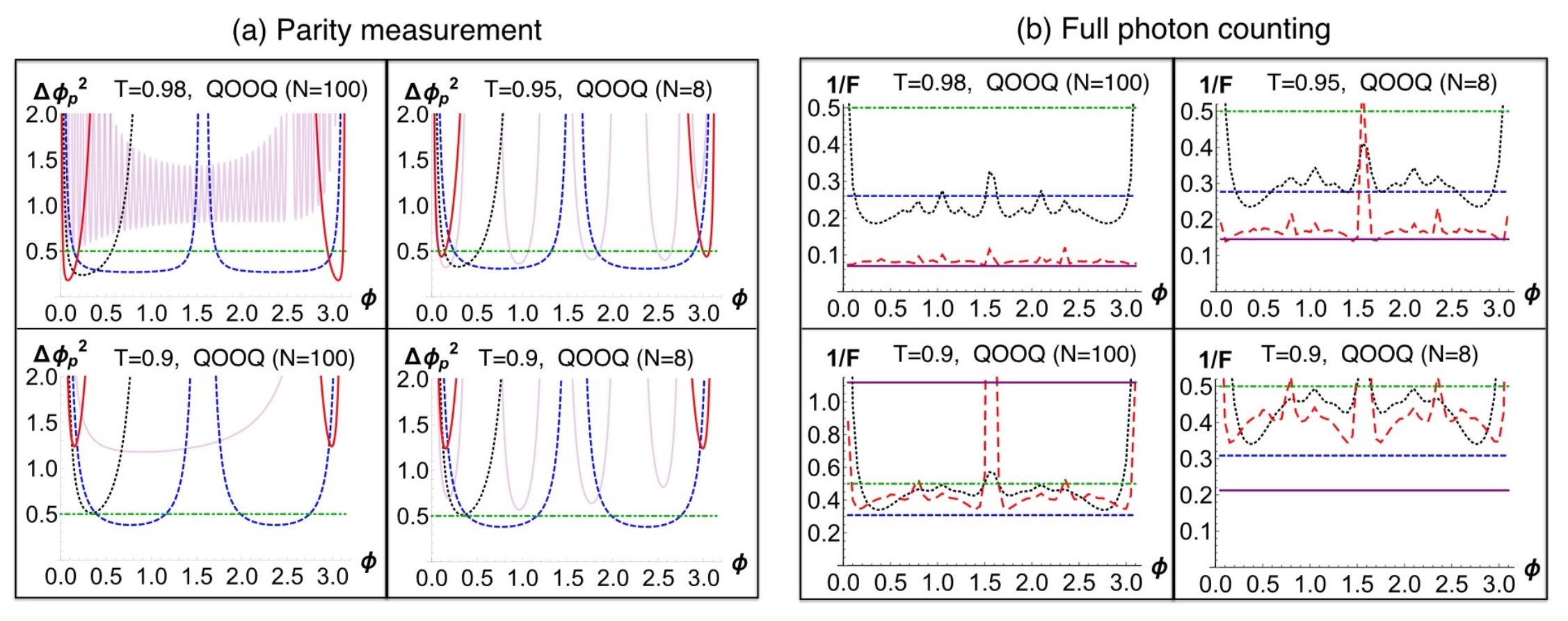}}}
  \caption{(a) Phase sensitivity $\Delta\phi^2_p$ achieved via parity measurement and (b) Phase sensitivity $1/F$ given by Fisher information $F$,  under a loss channel ($T$: transmittance rate) as a function of an unknown phase $\phi$, using $NOON$ (blue dashed curve), $AOOA$ (black dotted curve),
$SOOS$ (red long-dashed curve), and $QOOQ$ (purple solid curve) states. The shot-noise limit (green dashed line) is given by $(\Delta\phi_p)^2_{SNL}=1/N_{av}$. The average input photon number is set to be $N_{av}=2$. }
\label{fig:fig3}
\end{figure}

\subsection*{Measurement setup II: photon counting on both output modes}
As the second measurement scheme, we consider photon counting on both output modes in the MZI. The phase sensitivity is then given by classical Fisher information (FI) \cite{JD12}
\begin{eqnarray}
F=\sum_{n_a,n_b}\frac{1}{p(n_a,n_b|\phi)}\left[\frac{\partial p(n_a,n_b|\phi)}{\partial \phi}\right]^2,
\end{eqnarray}
where $p(n_a,n_b|\phi)$ is a probability of detecting $n_a$ photons on mode $a$ and $n_b$ photons on mode $b$ for a given phase $\phi$. For $m$ measurements, the FI provides Cram\'er-Rao bound (CRB), $(\delta\phi)^2\geq 1/(mF)$, and we are here interested in a single-shot measurement ($m=1$).

The detection probability for the state in Eq. (1) is given by
\begin{eqnarray}
p(n_a,n_b|\phi)=\frac{p_{n}2^{-n}}{1+p_0}\frac{n!}{n_a !n_b !}
[1+(-1)^{n_a}\cos(n\phi)],
\end{eqnarray}
where $n=n_a+n_b$. Remarkably, the corresponding Fisher information is the same as the quantum Fisher information of Eq. (3) and the path-symmetric entangled states achieve the optimal bound for the entire range $[0,2\pi]$ of unknown phase shifts in an ideal situation without loss.

On the other hand, when the state goes through loss in the two optical paths, the photon-counting probability is given by
\begin{eqnarray}
p(n_a,n_b|\phi)_T=\frac{p_{n}(T/2)^n}{1+p_0}\frac{n!}{n_a !n_b !}[1+(-1)^{n_a}\cos(n\phi)]
+\frac{1}{1+p_0} \left(\frac{T}{2}\right)^n\frac{1}{n_a !n_b !}\sum^{\infty}_{k=1}p_{n+k}\frac{(n+k)!}{k!}R^k,
\end{eqnarray}
where $n=n_a+n_b$ and $R=1-T$.
In Fig. 3 (b), we see that $QOOQ$ state with $N=8$ is more robust against the loss channel ($T=0.9$) than the other states over the entire range $[0,2\pi]$ of phase shifts. 
Typically, a high photon number state of a single-mode component is more vulnerable to the loss channel and the robustness of the path-symmetic entangled states shows the ordering $QOOQ(N=100)<AOOA<SOOS<NOON<QOOQ(N=8)$. 
They are ordered according to the size of SNL-beating range of phase-shift angle.
Compared with the parity measurement scheme under the same degree of loss, the path-symmetric entangled states appear less vulnerable in this second scheme based on the full photon-counting statistics.

Note that both $NOON$ and $QOOQ$ states show performance not only more robust against loss than other states but also independent of (insensitive to) phase shift.
In Fig. 4 (a), we show that $QOOQ$ state can provide a better resource against loss over the whole range of input energy. 
For each transmittance rate $T$, we optimize $N$ of $QOOQ$ states $\sqrt{q}|1\rangle+\sqrt{1-q}|N\rangle$ for the best phase sensitivity (red dotted curve).
We see that under the loss of considered range $T=0.85\sim 1$, $QOOQ$ state with $N=8$ is optimal or close to the optimal state. In particular, at $T=0.9$, $QOOQ$ state with $N=8$ is optimal and beats $NOON$ state over the range of the average input energy $N_{av}>1$. Thus, $QOOQ$ state with $N=8$ is more practical than other $QOOQ$ states.
Furthermore, in Fig. 4 (b), we show the result of phase sensitivity by randomly sampling single-mode component states in a form $|\varphi\rangle=\sum_{n=1}\sqrt{p_n}|n\rangle$, where we still find $QOOQ$ state with $N=8$ near optimal.


\begin{figure}
\centerline{\scalebox{0.2}{\includegraphics[angle=0]{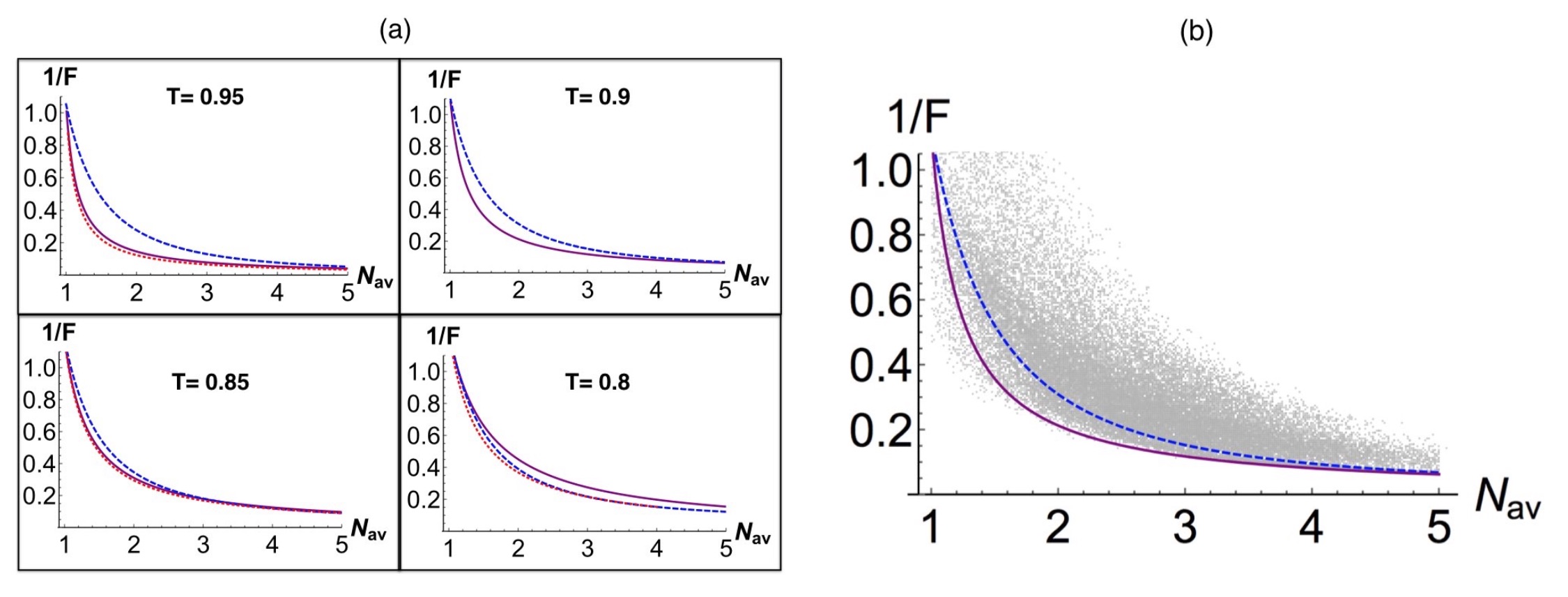}}}
  \caption{(a) Phase sensitivity $1/F$ given by Fisher information $F$ under a loss channel ($T$: transmittance rate) as a function of $N_{av}$, using $NOON$ state (blue dashed curve), $QOOQ$ state with $N=8$ (purple solid curve), and $QOOQ$ states with an optimized $N$ for each $T$ (red dotted curve).
At $T=0.95,~0.9,~0.85,~0.8$, the $QOOQ$ states are optimized with $N=13\sim25,~8\sim9,~5\sim6,~3\sim4$, respectively.~
(b) Phase sensitivity $1/F$ given by Fisher information $F$ under a loss channel ($T=0.9$) as a function of $N_{av}$, using $NOON$ state (blue dashed curve), $QOOQ$ state with $N=8$ (purple solid curve). In addition, random states in a form $\sum_{n=1}\sqrt{p_n}|n\rangle$ are sampled with a total number of $37,132$ (grey dots).
 }
\label{fig:fig4}
\end{figure}



\subsection*{State generation scheme}
In this section, we propose how to generate a resource state of Eq. (1) in experiment. The class of entangled states, $|\varphi\rangle_a|0\rangle_b+|0\rangle_a|\varphi\rangle_b$, may be generated by optical Fredkin gate \cite{GC01} utilizing a cross-Kerr nonlinearity and beam splitters. However, the cross-Kerr nonlinearity achieved in laboratory is rather limited, which can limit a successful generation of the resource states. Instead, we propose a scheme using a polarization entangled resource and conditional phase shift (CPS) operations. The CPS operation was implemented in a superconducting transmon qubit \cite{V13}. Specifically we are interested in the generation of $SOOS$ and $QOOQ$ states. The generation of high $NOON$ state ($N>4$) was proposed theoretically by the superposition of single (two)-photon operations \cite{KLD02,F02,LN12} or experimentally by post-selection after merging coherent laser light and down-converted photon \cite{HO07}. An $AOOA$ state may be generated by injecting a cat state and a coherent state into a beam splitter \cite{Joo11}.
However it is not known how to generate a $SOOS$ state or a $QOOQ$ state with currently available techniques. 

First, we propose a generation scheme for $SOOS$ state, which is implemented by a polarization entangled state $|H\rangle_u|V\rangle_d+|V\rangle_u|H\rangle_d$ and a conditional phase shift (CPS) operation $\hat{C}=I\otimes|H\rangle\langle H|+e^{ix\hat{a}^{\dag}\hat{a}}\otimes|V\rangle\langle V|$.
 Applying the CPS operation to target and ancillary qubits, the output qubits are given by
\begin{eqnarray}
\hat{C}_{a,u}\hat{C}_{b,d}|\xi\rangle_a|\xi\rangle_b\otimes(|H\rangle_u|V\rangle_d+|V\rangle_u|H\rangle_d)
=|\xi\rangle_a|\xi^{'}\rangle_b|H\rangle_u|V\rangle_d+|\xi^{'}\rangle_a|\xi\rangle_b|V\rangle_u|H\rangle_d,
\end{eqnarray}
where $\xi=re^{i\theta}$ and $\xi'=re^{i(\theta+2x)}$. Detecting the ancillary qubits to be in $|+\rangle=(|H\rangle+|V\rangle)/\sqrt{2}$, we obtain an entangled squeezed vacuum states, $|ESV\rangle=|\xi\rangle_a|\xi^{'}\rangle_b+|\xi^{'}\rangle_a|\xi\rangle_b$. Then, applying a single-mode squeezing operation on each mode, we obtain $SOOS$ state with $x=\pi/2$,
\begin{eqnarray}
\hat{S}(\xi)_a\hat{S}(\xi)_b|ESV\rangle=|2\xi\rangle_a|0\rangle_b+|0\rangle_a|2\xi\rangle_b.
\end{eqnarray}

In Fig. 5 (a), we show the setup for the $SOOS$ state. Two single-mode squeezed vacuum states $|\xi\rangle_a|\xi\rangle_b$ interact with ancillary qubits ($|H\rangle_u|V\rangle_d+|V\rangle_u|H\rangle_d$) under the CPS operation, where
the ancillary qubits are generated by spontaneous parametric down conversion in a nonlinear PPKTP crystal.
After the detection of the ancillary qubits, the $SOOS$ state is produced by the local operation $\hat{S}(\xi)$ on each output mode.
In general, when the single-mode component of the path-symmetric entangled state is decomposable with unitary operations, i.e. $|\varphi\rangle=\hat{U}^{\dag}e^{ix\hat{n}}\hat{U}|0\rangle$,  the generation scheme can be given by
\begin{eqnarray}
\frac{1}{\sqrt{2}}(|H\rangle_a|V\rangle_b+|V\rangle_a|H\rangle_b)\hat{U}_1\hat{U}_2|0\rangle_1|0\rangle_2
&\xrightarrow{CPS}& \frac{1}{\sqrt{2}}(\hat{U}_1|0\rangle_1e^{ix\hat{n}_2}\hat{U}_2|0\rangle_2+e^{ix\hat{n}_1}\hat{U}_1|0\rangle_1\hat{U}_2|0\rangle_2)\nonumber\\
&\xrightarrow{\hat{U}^{\dag}_1\hat{U}^{\dag}_2}& \frac{1}{\sqrt{2}}(|0\rangle_1\hat{U}^{\dag}_2e^{ix\hat{n}_2}\hat{U}_2|0\rangle_2+\hat{U}^{\dag}_1e^{ix\hat{n}_1}\hat{U}_1|0\rangle_1|0\rangle_2),
\end{eqnarray}
where $\hat{n}_{i}$ ($i=1,2$) is the photon number operator of mode $i$.

\begin{figure}
\centerline{\scalebox{0.2}{\includegraphics[angle=0]{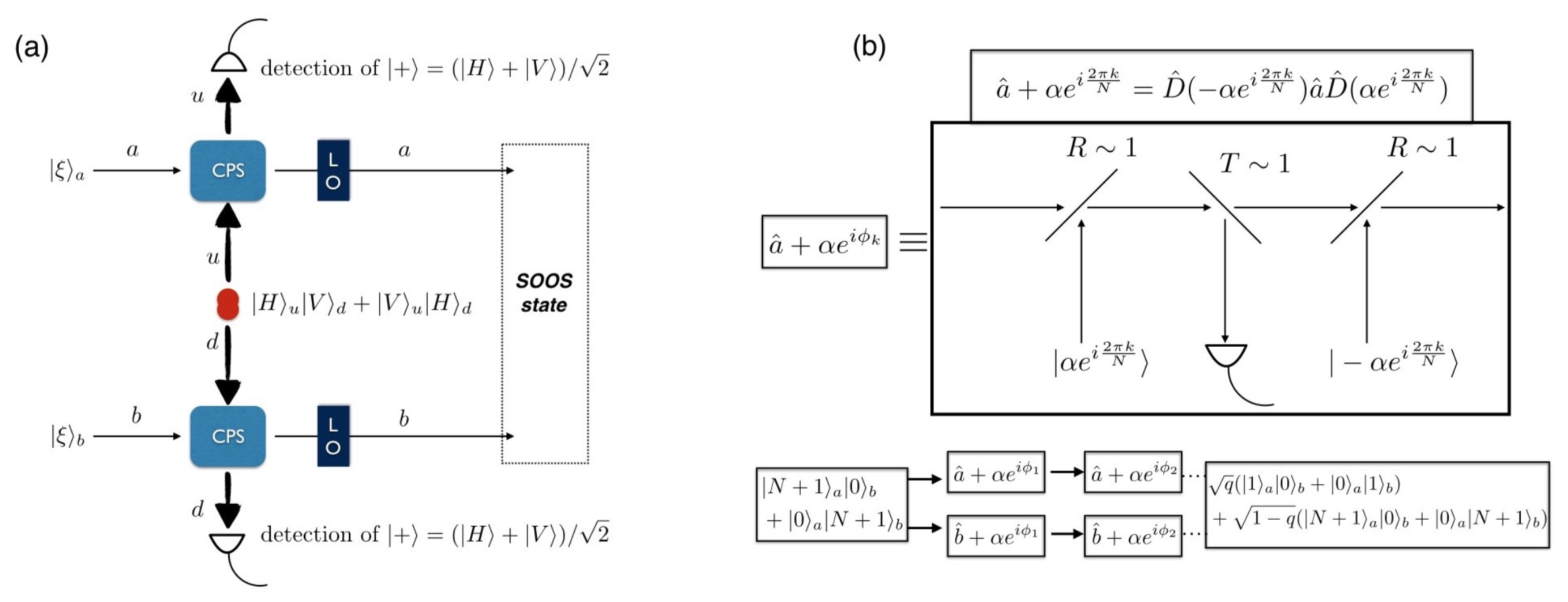}}}
  \caption{(a) Generation of $SOOS$ state via a conditional phase shift (CPS) operation. Each single-mode squeezed vacuum state is injected into a CPS operation setup. Ancillary qubits are entangled in polarization ($H,V$) and path ($u,d$) modes.
LO stands for local squeezing operation $\hat{S}(\xi)$.~
(b) Generation of $QOOQ$ state via a successive application of the coherent operation $\hat{a}+\alpha e^{i\phi_k}$ with $\phi_k=2\pi k/N$. $NOON$ state is first injected into the coherent operation setups. The coherent operation $\hat{a}+\alpha e^{i\phi_k}$ is implemented with three beam splitters. Two beam splitters are highly reflective whereas one beam splitter is highly transmissive. The phase $\phi_k$ is a control parameter of strong coherent light.
}
\label{fig:fig5}
\end{figure}


Second, we propose a generation scheme for $QOOQ$ state, which is obtained by a successive application of a coherent operation $\hat{a}+c$ with $c=\alpha e^{i2\pi k/N}$. Based on the transformation $\hat{a}+c=\hat{D}(-c)\hat{a}\hat{D}(c)$, the coherent operation is obtained by a sequential operation of displacement and photon subtraction operations. In Fig. 5 (b), we show that the coherent operation can be implemented with three beam splitters. The displacement operation is achieved with strong coherent light and a beam splitter with high reflectance \cite{LB02}. The photon subtraction operation is attained with a single-photon detection and a beam splitter with high transmittance \cite{WTG04}. Applying $N$ set of the coherent operations to an $N+1$ photon number state ($|N+1\rangle$), the output state is given by
\begin{eqnarray}
\prod^N_{k=1}(\hat{a}+\alpha e^{i\frac{2\pi k}{N}})|N+1\rangle
=\sqrt{(N+1)!}|1\rangle+(-1)^{N-1}\alpha^N|N+1\rangle.
\end{eqnarray}
For instance, initially preparing a $NOON$ state, $|N+1,0;0,N+1\rangle\equiv|N+1\rangle_a|0\rangle_b+|0\rangle_a|N+1\rangle_b$ 
and applying a set of coherent operations to each mode, as shown in Fig. 5 (b), we obtain $QOOQ$ state,
\begin{eqnarray}
\prod^N_{k=1}(\hat{a}+\alpha e^{i\frac{2\pi k}{N}})(\hat{b}+\alpha e^{i\frac{2\pi k}{N}})|N+1,0;0,N+1\rangle
=\sqrt{(N+1)!}|1,0;0,1\rangle+(-1)^{N-1}\alpha^N|N+1,0;0,N+1\rangle,
\end{eqnarray}
where $|1,0;0,1\rangle=|1\rangle_a|0\rangle_b+|0\rangle_a|1\rangle_b$.


\section*{Discussion}


We have introduced a generic class of path-symmetric entangled states $|\varphi\rangle|0\rangle+|0\rangle|\varphi\rangle$ for phase estimation, which includes the well-studied $NOON$ state and entangled-coherent states as special cases. The phase-sensitivity is largely determined by the photon statistics of the single-mode component state $|\varphi\rangle$, and in particular, the QFI is given in terms of the Mandel-$Q$ factor of $|\varphi\rangle$. 
We have shown that the QCRB can be lowered with the super-Poissonian statistics of $|\varphi\rangle$ and that even an arbitrarily small QCRB is possible with a finite energy, e.g. by a state of the form $|\varphi\rangle=\sqrt{q}|1\rangle+\sqrt{1-q}|N\rangle$.  On the other hand, for specific measurement schemes, we have considered a parity measurement and a full photon-counting method to obtain phase-sensitivity.  
Without photon loss, the latter scheme employing any path-symmetric states $|\varphi 0\rangle+|0\varphi\rangle$ achieves the QCRB over the entire range $[0,2\pi]$ of unknown phase shift $\phi$ whereas the former does so in a restricted range of $\phi$.
The case of $|\varphi\rangle=\sqrt{q}|1\rangle+\sqrt{1-q}|N\rangle$ turns out to provide the most robust resource against loss among the considered entangled states over the range of the average input energy $N_{av}>1$, particularly under the loss channel with transmittance rate $T\sim0.9$ currently available \cite{K10}.

Furthermore, we have proposed a generation scheme for $SOOS$ state by using polarization entangled states and conditional phase shift (CPS) operations, and a generation scheme for $QOOQ$ state by a successive application of a coherent operation $\hat{a}+c$. The generation scheme of the $SOOS$ state can also be modified to produce other type entangled states. Replacing one of the single-mode components with a coherent state before the CPS operation, we can produce a $SOOA$ state. Or, by controlling the phase shift operation and the local operations in the setup, we can produce $OOSS$ and $OOSA$ states.
Moreover, we may extend the squeezing component to the superposition of different squeezing operations which can provide more enhancement of QCRB, similar to the phenomenon by multi-headed cat state \cite{LLNK15}.
 
In this work we have addressed the estimation of a locally fixed, unknown, phase. A worthwhile direction for future work will be the estimation of a completely unknown phase over a finite range \cite{PS06,HW12}.
Moreover, we may investigate how quantum phase estimation can be improved with other entangled states by employing non-Gaussian operations, e.g., superposition of photon operations ($t\hat{a}+r\hat{a}^{\dag}$  and $t\hat{a}^2+r\hat{a}^{\dag 2}$)\cite{LN10,LN12} with a controllable parameter $t$ ($|t|^2+|r|^2=1$).

{\it Remarks}: 
While preparing this work, we became aware of a related work that also considered $SOOS$ state, differently termed as squeezed entangled state (SES) by Knott {\it et al.} in\cite{KCHPD15}. They showed that, in local quantum phase estimation without photon loss, the SES gives lower bound of phase sensitivity than $NOON$ state and a separable two-mode squeezed state $|\xi\rangle_a|\xi\rangle_b$. They also proposed a separable squeezed cat state (SCS), i.e. a product of two single-mode squeezed cat states, which provides even lower bound of the phase sensitivity than the SES. For a fixed average input energy $N_{av}=1$, the separable SCS exhibits advantage against $27\%$ loss in terms of QFI and $10\%$ loss under a full photon counting scheme.
In comparison to \cite{KCHPD15}, we also considered the $SOOS$ state equivalent to the SES, which actually belongs to a broader class of path-symmetric entangled states $|\varphi\rangle|0\rangle+|0\rangle|\varphi\rangle$ in our work. We showed that $QOOQ$ state among them provided much lower bound of phase sensitivity than the $SOOS$ state. We also proposed experimental schemes to generate those path-entangled states using current technology, though they could be practically demanding. Under the parity measurement scheme, we showed the sub-Poisonianity of the single-mode component in the states $|\varphi\rangle|0\rangle+|0\rangle|\varphi\rangle$ exhibits the robustness over photon loss. Under the full photon counting scheme, we showed that the $QOOQ$ state is more robust than any other states $|\varphi\rangle|0\rangle+|0\rangle|\varphi\rangle$, under $10\%$ loss, over the range of the average input energy $N_{av}> 1$.



\begin{thebibliography}{99}
  	
	\expandafter\ifx\csname url\endcsname\relax
	\def\url#1{\texttt{#1}}\fi
	\expandafter\ifx\csname urlprefix\endcsname\relax\def\urlprefix{DOI: }\fi
	\providecommand{\bibinfo}[2]{#2}
	\providecommand{\eprint}[2][]{\url{#2}}
	
	
	\bibitem{A16}
	\bibinfo{author}{Abbott, B. P.} \emph{et~al.}
	\newblock \bibinfo{title}{Observation of Gravitational Waves from a Binary Black Hole Merger}.
	\newblock \emph{\bibinfo{journal}{Phys. Rev. Lett.}} \textbf{\bibinfo{volume}{116}},
	\bibinfo{pages}{061102} ;
	\newblock \urlprefix{10.1103/PhysRevLett.116.061102} (\bibinfo{year}{2016}).





	\bibitem{Taylor13}
	\bibinfo{author}{Taylor, M.~A.} \emph{et~al.}
	\newblock \bibinfo{title}{Biological measurement beyond the quantum limit}.
	\newblock \emph{\bibinfo{journal}{Nat Photon}} \textbf{\bibinfo{volume}{7}},
	\bibinfo{pages}{229--233} ;
	\newblock \urlprefix{10.1038/nphoton.2012.346} (\bibinfo{year}{2013}).
	
	\bibitem{C81}
	\bibinfo{author}{Caves, C.~M.}
	\newblock \bibinfo{title}{Quantum-mechanical noise in an interferometer}.
	\newblock \emph{\bibinfo{journal}{Phys. Rev. D}} \textbf{\bibinfo{volume}{23}},
	\bibinfo{pages}{1693--1708}; 
	\newblock \urlprefix{10.1103/PhysRevD.23.1693} (\bibinfo{year}{1981}).
	
	
	\bibitem{GLM06}
	\bibinfo{author}{Giovannetti, V.}, \bibinfo{author}{Lloyd, S.} \&
	\bibinfo{author}{Maccone, L.}
	\newblock \bibinfo{title}{Quantum metrology}.
	\newblock \emph{\bibinfo{journal}{Phys. Rev. Lett.}}\textbf{\bibinfo{volume}{96}}, \bibinfo{pages}{010401};
	\newblock \urlprefix{10.1103/PhysRevLett.96.010401} (\bibinfo{year}{2006}).
	
	\bibitem{BC94}
	\bibinfo{author}{Braunstein, S.~L.} \& \bibinfo{author}{Caves, C.~M.}
	\newblock \bibinfo{title}{Statistical distance and the geometry of quantum states}.
	\newblock \emph{\bibinfo{journal}{Phys. Rev. Lett.}} \textbf{\bibinfo{volume}{72}}, \bibinfo{pages}{3439--3443};
	\newblock \urlprefix{10.1103/PhysRevLett.72.3439} (\bibinfo{year}{1994}).
	
	
	
	
	
	\bibitem{D08}
	\bibinfo{author}{Dowling, J.~P.}
	\newblock \bibinfo{title}{Quantum optical metrology -- the lowdown on high-n00n states}.
	\newblock \emph{\bibinfo{journal}{Contemporary Physics}}
	\textbf{\bibinfo{volume}{49}}, \bibinfo{pages}{125--143};
	\newblock \urlprefix{10.1080/00107510802091298} (\bibinfo{year}{2008}).

	

    \bibitem{Tilma10}
     \bibinfo{author}{Tilma, T.}, \bibinfo{author}{Hamaji, S.}, 
     \bibinfo{author}{Munro, W.~J.} \& \bibinfo{author}{Nemoto, K.}
 \newblock \bibinfo{title}{Entanglement is not a critical resource for quantum metrology}.
 \newblock \emph{\bibinfo{journal}{Phys. Rev. A}}
	\textbf{\bibinfo{volume}{81}}, \bibinfo{pages}{022108};
	\newblock \urlprefix{10.1103/PhysRevA.81.022108} (\bibinfo{year}{2010}).
	
	
	
	\bibitem{A10}
	\bibinfo{author}{Anisimov, P.~M.} \emph{et~al.}
	\newblock \bibinfo{title}{Quantum metrology with two-mode squeezed vacuum:
		Parity detection beats the heisenberg limit}.
	\newblock \emph{\bibinfo{journal}{Phys. Rev. Lett.}}
	\textbf{\bibinfo{volume}{104}}, \bibinfo{pages}{103602};
	\newblock \urlprefix{10.1103/PhysRevLett.104.103602} (\bibinfo{year}{2010}).


	
	\bibitem{GM10}
	\bibinfo{author}{Gerry, C.~C.} \& \bibinfo{author}{Mimih, J.}
	\newblock \bibinfo{title}{Heisenberg-limited interferometry with pair coherent
		states and parity measurements}.
	\newblock \emph{\bibinfo{journal}{Phys. Rev. A}} \textbf{\bibinfo{volume}{82}},
	\bibinfo{pages}{013831}; 
	\newblock \urlprefix{10.1103/PhysRevA.82.013831} (\bibinfo{year}{2010}).
	
	\bibitem{GLM11}
	\bibinfo{author}{Giovannetti, V.}, \bibinfo{author}{Lloyd, S.} \&
	\bibinfo{author}{Maccone, L.}
	\newblock \bibinfo{title}{Advances in quantum metrology}.
	\newblock \emph{\bibinfo{journal}{Nat Photon}} \textbf{\bibinfo{volume}{5}},
	\bibinfo{pages}{222--229}; 
	\newblock \urlprefix{10.1038/nphoton.2011.35} (\bibinfo{year}{2011}).
	
	\bibitem{BSV15}
	\bibinfo{author}{Bohmann, M.}, \bibinfo{author}{Sperling, J.} \&
	\bibinfo{author}{Vogel, W.}
	\newblock \bibinfo{title}{Entanglement and phase properties of noisy noon
		states}.
	\newblock \emph{\bibinfo{journal}{Phys. Rev. A}} \textbf{\bibinfo{volume}{91}},
	\bibinfo{pages}{042332} ;
	\newblock \urlprefix{10.1103/PhysRevA.91.042332} (\bibinfo{year}{2015}).
	
	\bibitem{GC01}
	\bibinfo{author}{Gerry, C.~C.} \& \bibinfo{author}{Campos, R.~A.}
	\newblock \bibinfo{title}{Generation of maximally entangled photonic states
		with a quantum-optical fredkin gate}.
	\newblock \emph{\bibinfo{journal}{Phys. Rev. A}} \textbf{\bibinfo{volume}{64}},
	\bibinfo{pages}{063814}; 
	\newblock \urlprefix{10.1103/PhysRevA.64.063814} (\bibinfo{year}{2001}).
	
	\bibitem{GB02}
	\bibinfo{author}{Gerry, C.~C.} \& \bibinfo{author}{Benmoussa, A.}
	\newblock \bibinfo{title}{Heisenberg-limited interferometry and
		photolithography with nonlinear four-wave mixing}.
	\newblock \emph{\bibinfo{journal}{Phys. Rev. A}} \textbf{\bibinfo{volume}{65}},
	\bibinfo{pages}{033822} ;
	\newblock \urlprefix{10.1103/PhysRevA.65.033822} (\bibinfo{year}{2002}).
	
	\bibitem{Joo11}
	\bibinfo{author}{Joo, J.}, \bibinfo{author}{Munro, W.~J.} \&
	\bibinfo{author}{Spiller, T.~P.}
	\newblock \bibinfo{title}{Quantum metrology with entangled coherent states}.
	\newblock \emph{\bibinfo{journal}{Phys. Rev. Lett.}}
	\textbf{\bibinfo{volume}{107}}, \bibinfo{pages}{083601};
	\newblock \urlprefix{10.1103/PhysRevLett.107.083601} (\bibinfo{year}{2011}).
	
	\bibitem{J12}
	\bibinfo{author}{Joo, J.} \emph{et~al.}
	\newblock \bibinfo{title}{Quantum metrology for nonlinear phase shifts with
		entangled coherent states}.
	\newblock \emph{\bibinfo{journal}{Phys. Rev. A}} \textbf{\bibinfo{volume}{86}},
	\bibinfo{pages}{043828}; 
	\newblock \urlprefix{10.1103/PhysRevA.86.043828} (\bibinfo{year}{2012}).
	
	\bibitem{H09}
	\bibinfo{author}{Hofmann, H.~F.}
	\newblock \bibinfo{title}{All path-symmetric pure states achieve their maximal
		phase sensitivity in conventional two-path interferometry}.
	\newblock \emph{\bibinfo{journal}{Phys. Rev. A}} \textbf{\bibinfo{volume}{79}},
	\bibinfo{pages}{033822} ;
	\newblock \urlprefix{10.1103/PhysRevA.79.033822} (\bibinfo{year}{2009}).
	
	\bibitem{SKDL13}
	\bibinfo{author}{Seshadreesan, K.~P.}, \bibinfo{author}{Kim, S.},
	\bibinfo{author}{Dowling, J.~P.} \& \bibinfo{author}{Lee, H.}
	\newblock \bibinfo{title}{Phase estimation at the quantum cram\'er-rao bound
		via parity detection}.
	\newblock \emph{\bibinfo{journal}{Phys. Rev. A}} \textbf{\bibinfo{volume}{87}},
	\bibinfo{pages}{043833} ;
	\newblock \urlprefix{10.1103/PhysRevA.87.043833} (\bibinfo{year}{2013}).
	
	
	\bibitem{RL12}
	\bibinfo{author}{Rivas, A.} \& \bibinfo{author}{Luis, A.}
	\newblock \bibinfo{title}{Sub-Heisenberg estimation of non-random phase shifts}.
	\newblock \emph{\bibinfo{journal}{New J. Phys.}} \textbf{\bibinfo{volume}{14}},
	\bibinfo{pages}{093052} ;
	\newblock \urlprefix{10.1088/1367-2630/14/9/093052} (\bibinfo{year}{2012}).
	
	
	\bibitem{Zhang13}
	\bibinfo{author}{Zhang, Y.~R.},  \bibinfo{author}{Jin, G.~R.},
	\bibinfo{author}{Cao, J.~P.}, \bibinfo{author}{Liu, W.~M.},
	\& \bibinfo{author}{Fan, H.}
	\newblock \bibinfo{title}{Unbounded quantum Fisher information in two-path interferometry with finite photon number}.
	\newblock \emph{\bibinfo{journal}{J. Phys. A}} \textbf{\bibinfo{volume}{46}},
	\bibinfo{pages}{035302} ;
	\newblock \urlprefix{10.1088/1751-8113/46/3/035302} (\bibinfo{year}{2013}).
	
			
	
	\bibitem{BIWH96}
	\bibinfo{author}{Bollinger, J. J.~.}, \bibinfo{author}{Itano, W.~M.},
	\bibinfo{author}{Wineland, D.~J.} \& \bibinfo{author}{Heinzen, D.~J.}
	\newblock \bibinfo{title}{Optimal frequency measurements with maximally
		correlated states}.
	\newblock \emph{\bibinfo{journal}{Phys. Rev. A}} \textbf{\bibinfo{volume}{54}},
	\bibinfo{pages}{R4649--R4652} ;
	\newblock \urlprefix{10.1103/PhysRevA.54.R4649} (\bibinfo{year}{1996}).
	
	\bibitem{C00}
	\bibinfo{author}{Gerry, C.~C.}
	\newblock \bibinfo{title}{Heisenberg-limit interferometry with four-wave mixers
		operating in a nonlinear regime}.
	\newblock \emph{\bibinfo{journal}{Phys. Rev. A}} \textbf{\bibinfo{volume}{61}},
	\bibinfo{pages}{043811} ;
	\newblock \urlprefix{10.1103/PhysRevA.61.043811} (\bibinfo{year}{2000}).
	
	\bibitem{CIDE14}
	\bibinfo{author}{Cohen, L.}, \bibinfo{author}{Istrati, D.},
	\bibinfo{author}{Dovrat, L.} \& \bibinfo{author}{Eisenberg, H.~S.}
	\newblock \bibinfo{title}{Super-resolved phase measurements at the shot noise
		limit by parity measurement}.
	\newblock \emph{\bibinfo{journal}{Opt. Express}} \textbf{\bibinfo{volume}{22}},
	\bibinfo{pages}{11945--11953} ;
	\newblock \urlprefix{10.1364/OE.22.011945} (\bibinfo{year}{2014}).
	
	\bibitem{DJA13}
	\bibinfo{author}{Distante, E.}, \bibinfo{author}{Je\ifmmode \check{z}\else \v{z}\fi{}ek, M.},
	\& \bibinfo{author}{Andersen, U. L.} 
	\newblock \bibinfo{title}{Deterministic Superresolution with Coherent States at the Shot Noise Limit}.
	\newblock \emph{\bibinfo{journal}{Phys. Rev. Lett.}} \textbf{\bibinfo{volume}{111}},
	\bibinfo{pages}{033603} ;
	\newblock \urlprefix{10.1103/PhysRevLett.111.033603} (\bibinfo{year}{2013}).
	
	
	
	\bibitem{K10}
	\bibinfo{author}{KacprowiczM.}, \bibinfo{author}{Demkowicz-DobrzanskiR.},
	\bibinfo{author}{WasilewskiW.}, \bibinfo{author}{BanaszekK.} \&
	\bibinfo{author}{A., W.}
	\newblock \bibinfo{title}{Experimental quantum-enhanced estimation of a lossy
		phase shift}.
	\newblock \emph{\bibinfo{journal}{Nat Photon}} \textbf{\bibinfo{volume}{4}},
	\bibinfo{pages}{357--360} ;
	\newblock \urlprefix{10.1038/nphoton.2010.39} (\bibinfo{year}{2010}).
	
	\bibitem{HBZW12}
	\bibinfo{author}{Hall, M.J.W.}, \bibinfo{author}{Berry, D.~W.},
	\bibinfo{author}{Zwierz, M.} \& \bibinfo{author}{Wiseman, H.~M.}
	\newblock \bibinfo{title}{Universality of the Heisenberg limit for estimates of random phase shifts}.
	\newblock \emph{\bibinfo{journal}{Phys. Rev. A}}
	\textbf{\bibinfo{volume}{85}}, \bibinfo{pages}{041802};
	\newblock \urlprefix{10.1103/PhysRevA.85.041802} (\bibinfo{year}{2012}).
	
	
	\bibitem{Hall12}
	\bibinfo{author}{Hall, M.J.W.}  \& \bibinfo{author}{Wiseman, H.~M.}
	\newblock \bibinfo{title}{Heisenberg-style bounds for arbitrary estimates of shift parameters including prior information}.
	\newblock \emph{\bibinfo{journal}{New J. Phys.}}
	\textbf{\bibinfo{volume}{14}}, \bibinfo{pages}{033040};
	\newblock \urlprefix{10.1088/1367-2630/14/3/033040} (\bibinfo{year}{2012}).
	
	
	
	\bibitem{SQ15}
	\bibinfo{author}{Sahota, J.} \& \bibinfo{author}{Quesada, N.}
	\newblock \bibinfo{title}{Quantum correlations in optical metrology:
		Heisenberg-limited phase estimation without mode entanglement}.
	\newblock \emph{\bibinfo{journal}{Phys. Rev. A}} \textbf{\bibinfo{volume}{91}},
	\bibinfo{pages}{013808} ;
	\newblock \urlprefix{10.1103/PhysRevA.91.013808} (\bibinfo{year}{2015}).
	
	
	\bibitem{BCM96}
	\bibinfo{author}{Braunstein, S.L.},  \bibinfo{author}{Caves, C.M.},
	\& \bibinfo{author}{Milburn, G.J.}
	\newblock \bibinfo{title}{Generalized Uncertainty Relations: Theory, Examples, and Lorentz Invariance}.
	\newblock \emph{\bibinfo{journal}{Ann. Phys.}} \textbf{\bibinfo{volume}{247}},
	\bibinfo{pages}{135 - 173} ;
	\newblock \urlprefix{10.1006/aphy.1996.0040} (\bibinfo{year}{1996}).
	
	
	
	
	
	\bibitem{JD12}
	\bibinfo{author}{Jarzyna, M.} \&
	\bibinfo{author}{Demkowicz-Dobrza\ifmmode~\acute{n}\else \'{n}\fi{}ski, R.}
	\newblock \bibinfo{title}{Quantum interferometry with and without an external
		phase reference}.
	\newblock \emph{\bibinfo{journal}{Phys. Rev. A}} \textbf{\bibinfo{volume}{85}},
	\bibinfo{pages}{011801} ;
	\newblock \urlprefix{10.1103/PhysRevA.85.011801} (\bibinfo{year}{2012}).
	
	\bibitem{ZPK10}
	\bibinfo{author}{Zwierz, M.}, \bibinfo{author}{P\'erez-Delgado, C.~A.} \&
	\bibinfo{author}{Kok, P.}
	\newblock \bibinfo{title}{General optimality of the heisenberg limit for
		quantum metrology}.
	\newblock \emph{\bibinfo{journal}{Phys. Rev. Lett.}}
	\textbf{\bibinfo{volume}{105}}, \bibinfo{pages}{180402};
	\newblock \urlprefix{10.1103/PhysRevLett.105.180402} (\bibinfo{year}{2010}).
	
	\bibitem{KSD11}
	\bibinfo{author}{Knysh, S.}, \bibinfo{author}{Smelyanskiy, V.~N.} \&
	\bibinfo{author}{Durkin, G.~A.}
	\newblock \bibinfo{title}{Scaling laws for precision in quantum interferometry
		and the bifurcation landscape of the optimal state}.
	\newblock \emph{\bibinfo{journal}{Phys. Rev. A}} \textbf{\bibinfo{volume}{83}},
	\bibinfo{pages}{021804} ;
	\newblock \urlprefix{10.1103/PhysRevA.83.021804} (\bibinfo{year}{2011}).
	

	
	\bibitem{V13}
	\bibinfo{author}{Vlastakis, B.} \emph{et~al.}
	\newblock \bibinfo{title}{Deterministically encoding quantum information using
		100-photon schr{\"o}dinger cat states}.
	\newblock \emph{\bibinfo{journal}{Science}} \textbf{\bibinfo{volume}{342}},
	\bibinfo{pages}{607--610} ;
	\newblock \urlprefix{10.1126/science.1243289} (\bibinfo{year}{2013}).
	
	
	\bibitem{KLD02}
	\bibinfo{author}{Kok, P.}, \bibinfo{author}{Lee, H.} \&
	\bibinfo{author}{Dowling, J.~P.}
	\newblock \bibinfo{title}{Creation of large-photon-number path entanglement
		conditioned on photodetection}.
	\newblock \emph{\bibinfo{journal}{Phys. Rev. A}} \textbf{\bibinfo{volume}{65}},
	\bibinfo{pages}{052104} ;
	\newblock \urlprefix{10.1103/PhysRevA.65.052104} (\bibinfo{year}{2002}).
	
	\bibitem{F02}
	\bibinfo{author}{Fiur\'a\ifmmode~\check{s}\else \v{s}\fi{}ek, J.}
	\newblock \bibinfo{title}{Conditional generation of \textit{N} -photon
		entangled states of light}.
	\newblock \emph{\bibinfo{journal}{Phys. Rev. A}} \textbf{\bibinfo{volume}{65}},
	\bibinfo{pages}{053818} ;
	\newblock \urlprefix{10.1103/PhysRevA.65.053818} (\bibinfo{year}{2002}).
	
	\bibitem{LN12}
	\bibinfo{author}{Lee, S.-Y.} \& \bibinfo{author}{Nha, H.}
	\newblock \bibinfo{title}{Second-order superposition operations via
		hong-ou-mandel interference}.
	\newblock \emph{\bibinfo{journal}{Phys. Rev. A}} \textbf{\bibinfo{volume}{85}},
	\bibinfo{pages}{043816} ;
	\newblock \urlprefix{10.1103/PhysRevA.85.043816} (\bibinfo{year}{2012}).
	
	\bibitem{HO07}
	\bibinfo{author}{Hofmann, H.~F.} \& \bibinfo{author}{Ono, T.}
	\newblock \bibinfo{title}{High-photon-number path entanglement in the
		interference of spontaneously down-converted photon pairs with coherent laser
		light}.
	\newblock \emph{\bibinfo{journal}{Phys. Rev. A}} \textbf{\bibinfo{volume}{76}},
	\bibinfo{pages}{031806} ;
	\newblock \urlprefix{10.1103/PhysRevA.76.031806} (\bibinfo{year}{2007}).
	
	\bibitem{LB02}
	\bibinfo{author}{Lvovsky, A.~I.} \& \bibinfo{author}{Babichev, S.~A.}
	\newblock \bibinfo{title}{Synthesis and tomographic characterization of the
		displaced fock state of light}.
	\newblock \emph{\bibinfo{journal}{Phys. Rev. A}} \textbf{\bibinfo{volume}{66}},
	\bibinfo{pages}{011801} ;
	\newblock \urlprefix{10.1103/PhysRevA.66.011801} (\bibinfo{year}{2002}).
	
	\bibitem{WTG04}
	\bibinfo{author}{Wenger, J.}, \bibinfo{author}{Tualle-Brouri, R.} \&
	\bibinfo{author}{Grangier, P.}
	\newblock \bibinfo{title}{Non-gaussian statistics from individual pulses of
		squeezed light}.
	\newblock \emph{\bibinfo{journal}{Phys. Rev. Lett.}}
	\textbf{\bibinfo{volume}{92}}, \bibinfo{pages}{153601};
	\newblock \urlprefix{10.1103/PhysRevLett.92.153601} (\bibinfo{year}{2004}).
	
	\bibitem{LLNK15}
	\bibinfo{author}{Lee, S.-Y.}, \bibinfo{author}{Lee, C.-W.},
	\bibinfo{author}{Nha, H.} \& \bibinfo{author}{Kaszlikowski, D.}
	\newblock \bibinfo{title}{Quantum phase estimation using a multi-headed cat
		state}.
	\newblock \emph{\bibinfo{journal}{J. Opt. Soc. Am. B}}
	\textbf{\bibinfo{volume}{32}}, \bibinfo{pages}{1186--1192};
	\newblock \urlprefix{10.1364/JOSAB.32.001186} (\bibinfo{year}{2015}).
	
	\bibitem{PS06}
	\bibinfo{author}{Pezz\'e, L.} \& \bibinfo{author}{Smerzi, A.}
	\newblock \bibinfo{title}{Phase sensitivity of a mach-zehnder interferometer}.
	\newblock \emph{\bibinfo{journal}{Phys. Rev. A}} \textbf{\bibinfo{volume}{73}},
	\bibinfo{pages}{011801} ;
	\newblock \urlprefix{10.1103/PhysRevA.73.011801} (\bibinfo{year}{2006}).
	
	\bibitem{HW12}
	\bibinfo{author}{Hall, M. J.~W.} \& \bibinfo{author}{Wiseman, H.~M.}
	\newblock \bibinfo{title}{Does nonlinear metrology offer improved resolution?
		answers from quantum information theory}.
	\newblock \emph{\bibinfo{journal}{Phys. Rev. X}} \textbf{\bibinfo{volume}{2}},
	\bibinfo{pages}{041006} ;
	\newblock \urlprefix{10.1103/PhysRevX.2.041006} (\bibinfo{year}{2012}).
	
	\bibitem{LN10}
	\bibinfo{author}{Lee, S.-Y.} \& \bibinfo{author}{Nha, H.}
	\newblock \bibinfo{title}{Quantum state engineering by a coherent superposition
		of photon subtraction and addition}.
	\newblock \emph{\bibinfo{journal}{Phys. Rev. A}} \textbf{\bibinfo{volume}{82}},
	\bibinfo{pages}{053812} ;
	\newblock \urlprefix{10.1103/PhysRevA.82.053812} (\bibinfo{year}{2010}).
	
	\bibitem{KCHPD15}
	\bibinfo{author}{Knott, P.~A.}, \bibinfo{author}{Cooling, J.~P.},
	\bibinfo{author}{Hayes, A.}, \bibinfo{author}{Proctor, T.~J.} \&
	\bibinfo{author}{Dunningham, J.~A.}
	\newblock \bibinfo{title}{{Practical quantum metrology with large precision gains in the low-photon-number regime}}.
	\newblock \emph{\bibinfo{journal}{Phys. Rev. A}} \textbf{\bibinfo{volume}{93}},
	\bibinfo{pages}{033859};
	\newblock \urlprefix{10.1103/PhysRevA.93.033859} (\bibinfo{year}{2016}).
		
	
	
	
	
	
\end{thebibliography}

\section*{Acknowledgements (not compulsory)}
Authors greatly appreciate M.J.W. Hall and C.C. Gerry for useful comments.
This research was supported by the IT R$\&$D program of MOTIE/KEIT [$1004346~ (2012)$] and by an NPRP grant 7-210-1-032 from Qatar National Research Fund.


\section*{Author contributions statement}
S.-Y.L. developed the scenario with H.N. All authors have contributed to analyzing the results. S.-Y.L. wrote the manuscript and all authors reviewed.


\section*{Additional information}

To include, in this order: \textbf{Accession codes} (where applicable); \textbf{Competing financial interests} (The authors declare no competing financial interests.). 





\end{document}